# Lasing without population inversion in $N_2^+$

A. Mysyrowicz[1], R. Danylo[1,2], A. Houard[1], V. Tikhonchuk[3,4], X. Zhang[2], Z. Fan[2], Q. Liang[2], S. Zhuang[2], L. Yuan[5], Y. Liu[2,1,*]

[1]*Laboratoire d'Optique Appliquée, ENSTA, Ecole Polytechnique, CNRS, Université Paris-Saclay, 828 Boulevard des Maréchaux, 91762 Palaiseau cedex, France*
[2]*Shanghai Key Lab of Modern Optical System, University of Shanghai for Science and Technology, 516, Jungong Road, 200093 Shanghai, China*
[3]*Centre Lasers Intenses et Applications, University of Bordeaux-CNRS-CEA, 351 Cours de la Liberation, 33405 Talence cedex, France*
[4]*ELI-Beamlines, Institute of Physics, Czech Academy of Sciences, 25241 Dolní Břežany, Czech Republic*
[5]*School of Physics and Astronomy, Shanghai Jiao Tong University, Shanghai 200240, China*
[*]*Corresponding author:* * yi.liu@usst.edu.cn*

## Abstract

We demonstrate experimentally the existence of long-lived coherent polarizations coupling simultaneously ground state X ($X^2\Sigma_g^+$) to excited states A ($A^2\Pi_u$) and B ($B^2\Sigma_u^+$) of $N_2^+$ inside a plasma created by a short intense laser pulse at 800 nm. This three-level V scheme arrangement is responsible for a strong optical gain without population inversion at the B – X transition at 391.4 nm. Simulations based on Maxwell-Bloch equations reproduce well the kinetics and the pressure dependence of the gain.

## I. INTRODUCTION

When two electronic levels |1⟩ and |2⟩ in an atomic or molecular system are simultaneously coupled by two coherent light fields to a common level |3⟩, the corresponding amplitudes of transition probability can interfere. This leads to spectacular and counterintuitive effects such as electromagnetically induced transparency[1], the disappearance of spontaneous emission[2,3], or the possibility of amplification without population inversion[4,5]. While the first two effects are well documented, there are few experiments so far demonstrating lasing without population inversion[6-9]. In this work, we report on the occurrence of such a lasing without population inversion in a weakly ionized gas of nitrogen molecules.

The underdense plasma is created by an intense near infrared femtosecond laser pulse. It shows remarkable cavity-free forward lasing at 391.4 or 427.8 nm[10-21]. These wavelengths correspond to transitions between the excited state B ($B^2\Sigma_u^+$, 0) and ground state X($X^2\Sigma_g^+$, 0, 1) of the singly ionized nitrogen molecule, where index 0, 1 refer to the vibrational levels. Despite a considerable amount of publications, interpretation of this cavity-free lasing is still controversial. Several noteworthy effects have been observed at low pressures. A femtosecond seed pulse at wavelength 391.4 or 427.8 nm crossing a 1 cm long plasma produced by an 800 nm femtosecond pump pulse experiences an increase of energy by two orders of magnitude[14,16]. Measurements of the amplified emission as a function of delay $\Delta t$ between the pump and the UV seed pulse show that the onset of amplification occurs at $\Delta t = 0$ and then decreases with a time constant of ~ 10 ps[10,14]. Most surprisingly, time-resolved measurements of the 391.4 nm amplified emission at fixed delay $\Delta t = 1$ ps show that the seed pulse itself is practically unaffected when crossing the plasma. Instead, it triggers a **retarded** emission at the same wavelength, with a delay that decreases with increasing gas pressure[12,14]. Retarded emission at 391.4 nm, although weaker, is observed even in the absence of any seed pulse. In this case, the forward emission intensity shows a remarkable sensitivity to the pump pulse duration and wavelength[17]: rapid cyclic variations are observed in a small range of tuning of the pump pulse wavelength around 800 nm, with a cycle that depends on pulse duration. Finally, the forward emission intensity at 391.4 nm (measured in the absence of a seed) depends on the pump pulse ellipticity. Starting with a linearly polarized pump, it first slightly increases with ellipticity $\varepsilon$ until it reaches a maximum around $\varepsilon \sim 0.2$, at which point it decreases and disappears above $\varepsilon \sim 0.4$[14].

Two aspects of the results are particularly challenging to explain. First, how is optical amplification between levels B and X obtained under the given pumping conditions? Second, how can one explain a delayed amplified emission? Two categories of theoretical models or interpretations have been proposed but do not address these two questions simultaneously. One type of models concentrates on an explanation of amplification based on population inversion between B and X but ignore the delayed aspect of amplification and its pressure dependence. Either inversion of electronic population due to storage of population in intermediate level A[15, 16] or transient inversion of rotational populations without inversion of electronic populations[19] has been proposed. Another interpretation attributes the delayed emission to superradiance but fail to explain the origin of optical amplification[12, 14, 22].

**II. RESULTS AND DISCUSSION**

We show in this paper that it is possible to give a consistent interpretation of the experimental facts described above by introducing a three level (V-scheme) arrangement leading to lasing without population inversion. Figure 1 shows how such a V scheme is realized in $N_2^+$. To validate such a V-scheme it is necessary to demonstrate the presence of long-lived coherent polarizations coupling quasi-resonantly state X to state B and intermediate state A ($A^2\Pi_u$) of the molecular ion after the end of the pump pulse. We have obtained experimental evidence for these coherent couplings lasting for the duration of the gain in the following manner: we injected a second 10 times weaker femtosecond pulse at 800 nm after the main 800 nm pump pulse, and measured the intensity of the forward gain at 391.4 nm as a function of the delay between the two input pulses. The experimental setup is schematically presented in Fig. 2 (a). A pronounced modulation of the signal was observed by fine tuning the delay around several fixed values up to 5 ps. For instance, Figure 3 (a) shows the modulation pattern obtained by fine scanning the delay around 3 ps. The Fourier transform of the modulation yields a wavelength of 780.5 nm shown in Fig. 3 (c), close to the R branch of the X(0)-A(2) transition at 782.6 nm (see Fig. 1 (b)). Since the time separation between the two laser pulses largely exceeds their duration, the interference pattern provides evidence of a long lasting coherent polarization induced by the first pulse. A similar experiment involving two seed pulses at 391 nm (see setup in Fig. 2(b)) reveals the existence of a long-lived coherent polarization at 391.4 nm, corresponding to the B(0)-X(0) transition, as also shown in Fig. 3 (b) and (d) [23].

The theoretical model of amplified emission at 391.4 nm considers three levels of $N_2^+$ molecules, X(0), A(2) and B(0), coupled by two electromagnetic fields corresponding to X-A and X-B transitions. Following the approach of Svidzinsky et al. [5], we have calculated the signal amplification of a seed pulse at 391.4 nm by solving Maxwell-Bloch equations for the three level system B-X-A in the envelope approximation. This modeling is different from the previous theoretical considerations[15, 16] as it considers amplification times hundred times longer than the pump pulse, centimeter-size amplification lengths and no population inversion between the B and X states is assumed. The equations for the electric field amplitudes $E_{ax}$ and $E_{bx}$, the polarization amplitudes $\rho_{ax}$, $\rho_{bx}$, $\rho_{ba}$, and the populations $n_a$, $n_b$ and $n_x$ corresponding to the A-X and B-X transitions read:

$$c\partial_z \Omega_{ax} = i\frac{\omega_{ax} N_i \mu_{ax}^2}{\hbar \epsilon_0}\rho_{ax}; \quad \partial_\tau \rho_{ax} = -\gamma_{ax}\rho_{ax} - \frac{i}{2}(n_a - n_x)\Omega_{ax} - \frac{i}{2}\rho_{ba}^*\Omega_{bx}, \qquad (1)$$

$$c\partial_z \Omega_{bx} = i\frac{\omega_{bx} N_i \mu_{bx}^2}{\hbar \epsilon_0}\rho_{bx}; \quad \partial_\tau \rho_{bx} = -\gamma_{bx}\rho_{bx} - \frac{i}{2}(n_b - n_x)\Omega_{bx} - \frac{i}{2}\rho_{ba}\Omega_{ax}, \qquad (2)$$

$$\partial_\tau n_a = -\mathrm{Im}(\rho_{ax}^*\Omega_{ax}); \quad \partial_\tau n_b = -\mathrm{Im}(\rho_{bx}^*\Omega_{bx}); \quad \partial_\tau \rho_{ba} = -\gamma_{ba}\rho_{ba} + \frac{i}{2}\rho_{ax}^*\Omega_{bx} - \frac{i}{2}\rho_{bx}\Omega_{ax}^*.$$

Here the terms $\gamma_{ij}$ describe the spontaneous decay rates of the excited states, $\Omega_{ax,bx} = \mu_{ax,bx}E_{ax,bx}/\hbar$ are the corresponding Rabi frequencies, $\tau = t - z/c$ is the co-propagation time, $\hbar$ is the Planck constant, $\epsilon_0$ is the dielectric permittivity of vacuum, $\mu_{ax}$ and $\mu_{bx}$ are the dipole moments corresponding to A-X and B-X transitions [24, 25] and $N_i$ is the density of molecular ions. The last terms in Eqs. (1) and (2) account for the quantum interference between states A and B via the cross-polarization $\rho_{ba}$. The two following conditions are assumed in the simulations: (i) 100 fs after the peak of the main laser pulse (considered as $t = 0$ in the simulation), the system contains 10% of $N_2^+$ ions, which are distributed between the ground and exited states as follows: a relative population of 60% in the ground state X, 30% in the excited state B, and 10% in the excited state A; (ii) the measured long-lived polarization at 780.5 nm is maintained by the decaying tail of the pump pulse slightly detuned from the R branch of the A(2) - X(0) transition at 782.6 nm.

The choice of distribution of ions between three levels is motivated by analysis of excitation probabilities of $N_2^+$ ions by the main laser pulse. The presence of a weak exponential post pulse decaying on a time scale of 5 ps was confirmed by measurements with a high contrast cross-correlator shown in Fig. 4. Such a post-pulse is common in Ti: Sapphire laser systems based on chirped pulse amplification technology[26]. This tail is approximated in the simulations by an exponential decay with a time constant $\tau_p$ = 5 ps and an initial intensity ~ 30 GW/cm$^2$. A short and weak seed pulse at 391 nm arrives 0.5 ps after the main pulse. Its intensity is thousand times smaller than the post-pulse intensity and its duration is 0.1 ps. While varying gas pressure we assume a constant ionization at the level of 10%.

Results of simulations, shown in Fig. 5, indicate that the seed pulse at 391.4 nm triggers a strong delayed amplified emission at the same wavelength provided condition $n_X > n_B > n_A$ is satisfied. That amplification is due to parametric coupling between the B and A level by the cross-polarization term in eqs. (1) and (2). It is also known in the literature under the name of *two-photon Raman effect*[27, 28]. Following Ref. 5-7 we prefer to call it *lasing without population inversion* to stress the fact that the initial state of the amplified transition B is less populated than the final state X. The measured and calculated amplified pulse in Fig. 5 show good agreement in the temporal shape and duration.

A better insight in the amplification process could be gained by considering evolution of populations in the states A, B and X displayed in Fig. 6 along with the intensity of 391.4 nm emission. One can see relaxation oscillations between levels A and X at a Rabi frequency, increase of the population at the level A and the correspondent decrease of the population at the levels B and X. During the whole process, the population of B state is less than 30%, while that of X state varies between 60% and 35%. We noticed that no population inversion between B and X occurs during the whole lasing process. The parametric coupling between B and A states leads to an exponential signal growth at 391 nm with a characteristic time of a few picoseconds. The growth rate of the signal is proportional to the population difference between levels B and A and the amplitude of the laser post-pulse. Amplification proceeds efficiently if the detuning of the long lasting polarization from the A-X resonance is smaller than the corresponding Rabi frequency.

Besides reproducing the delayed amplification and the pulse shape, the simulations restitute the measured dependence of lasing on gas pressure in the range *p* = 5-100 mbar, as shown in Fig. 6. In the current simulations, we neglected the possible variation of pump laser intensity and plasma length for different gas pressures, in order to keep the simplicity of the calculation. This can be the reason for the slight discrepancy in the optimum pressure between experiments and simulations in Fig. 7. As will be discussed in a more detailed forthcoming paper, the increase of gain at pressures lower than 30-40 mbar is explained by increasing number of excited ions. The reduction of gain at higher pressures is due to a post-pulse-induced transition from the level X to the level A. That process is progressively depleting the pump and destroying the amplification condition $n_B > n_A$.

Other experimental facts support the interpretation of amplification based on the three-level coupling without population inversion between B and X. Gain is expected to disappear if the pump wavelength is largely detuned from the required A-X resonance. Indeed, no amplification is found for a pump pulse tuned to the mid IR, except if the pump laser wavelength corresponds to a sub multiple of the B-X transition. In that case, enhanced emission at 391.4 nm is observed, but with different characteristics: the enhanced emission starts immediately without delay and no amplification upon injection of a seed pulse is observed. This mid IR-induced emission at 391.4 nm has been attributed to resonantly-enhanced third and fifth harmonics[29]. On the other hand, the V-scheme predicts the possibility of lasing without population inversion using a short pump pulse tuned around 1.1 μm. The V-coupling would then correspond to the B(0)-X(0)-A(0) scheme shown in Fig. 2 (b).

The model of three levels coupled by a laser post-pulse also sheds light on the origin of gain in the absence of a seed pulse that was reported in Ref. 17. In that case, the X-B coherence normally created by the seed pulse is replaced by a coherent polarization induced by recollisions from electrons driven by the pump pulse field, at twice the pump frequency. Also, the strong decrease of the gain upon increasing pump pulse ellipticity (observed in the absence of a seed pulse) derives naturally from the recollision model [30, 31], since with circularly polarized pump light, free electron trajectories miss the molecular core, preventing recollisions and therefore suppressing the required coherent polarization between X and B states.

At this point we wish to stress that our V-coupling model applied to $N_2^+$ is still incomplete. Rotation of the nitrogen molecular ions has to be accounted for. As discussed in Refs. 18-20, gain may be periodically enhanced due to laser-induced rotation of molecules. This can lead to a considerable increase of the signal at specific delays determined by the revivals of rotational packets.

## III. CONCLUSION

In conclusion, we have presented evidence for long-lasting coherences for AX and BX transitions. A three-level model accounting for the coupling between those coherences predicts parametric amplification of the BX transition without population inversion between these two states. It allows solving a longstanding puzzle, the emergence of an important delayed optical gain in a gas of nitrogen molecules or air pumped by a short intense laser pulse at 800 nm. This provides one of the few known examples of lasing without population inversion. It could occur in other molecular or atomic gases when the frequency of a short and intense pump pulse is in resonance with a transition between levels of ionized atoms or molecules.

**Acknowledgments**

The work is supported by the National Natural Science Foundation of China (Grants No. 11574213), Innovation Program of Shanghai Municipal Education Commission (Grant No. 2017-01-07-00-07-E00007), ELITAS (ELI Tools for Advanced Simulation) CZ.02.1.01/0.0/0.0/16_013/0001793 from the European Regional Development Fund. The authors acknowledge Dr. xx, Dr. xx, and Prof. Yi Xu of Shanghai Institute of Optics and Fine Mechanics for their technical support and help for the laser contrast measurement.

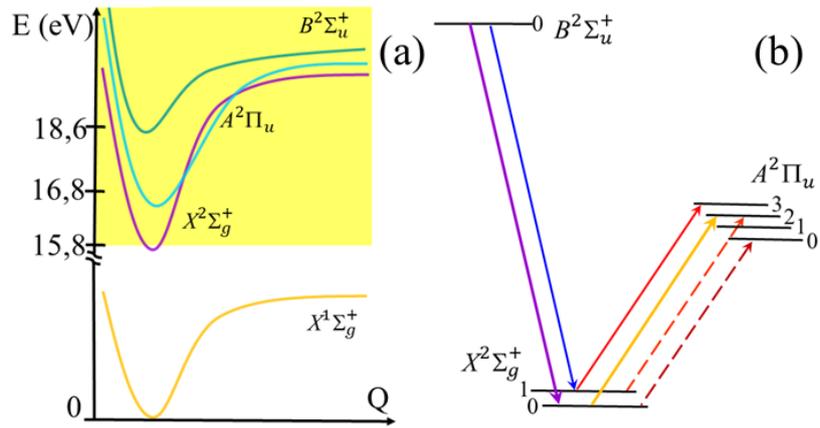

Fig. 1. (a): Schematic representation of relevant energy levels of nitrogen molecules in function of internuclear distance $Q$: $X^1\Sigma_g^+$ refers to the ground state of the neutral molecule, $X^2\Sigma_g^+$, $A^2\Pi_u$, $B^2\Sigma_u^+$ to the first three levels of the singly ionized molecule. (b): Schematic representation of the V-scheme. Ascending arrows represent the pump pulse, descending arrows the emitted pulse.

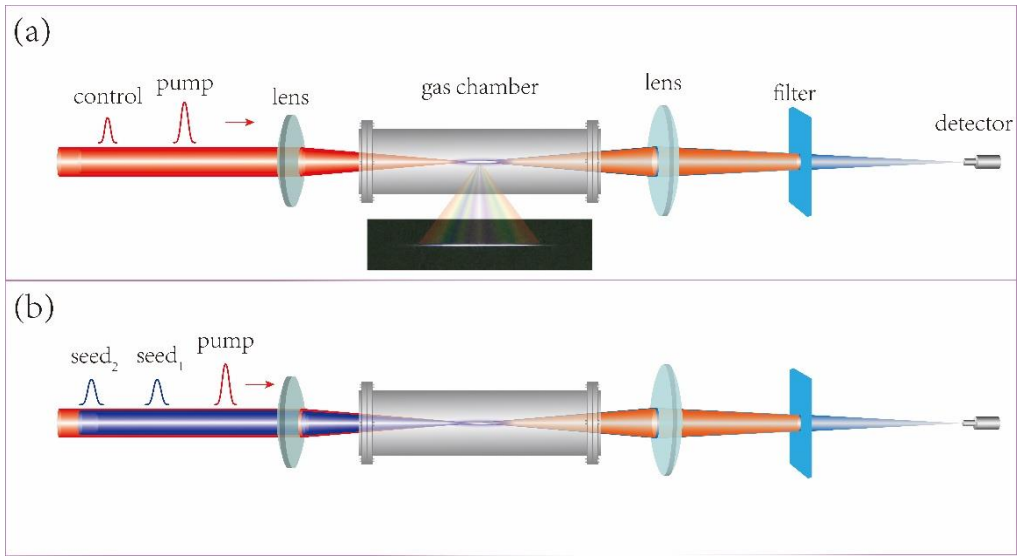

Fig. 2. (a), Two successive pulses at 800 nm with typical energy of 2 and 0.1 mJ and 40 fs pulse duration are focused with a $f = 40$ cm lens in a chamber filled with nitrogen gas at a pressure of 8 mbar. (b), Two seeding pulses around 391 nm with an adjustable time delay are injected into the nitrogen gas plasma formed by the 800 nm pump pulse. The output signal around 391 nm is detected as a function of fine tuning of time delay between the two 800 nm pulses in (a), and as a function of the time delay between the two seeding pulses in (b).

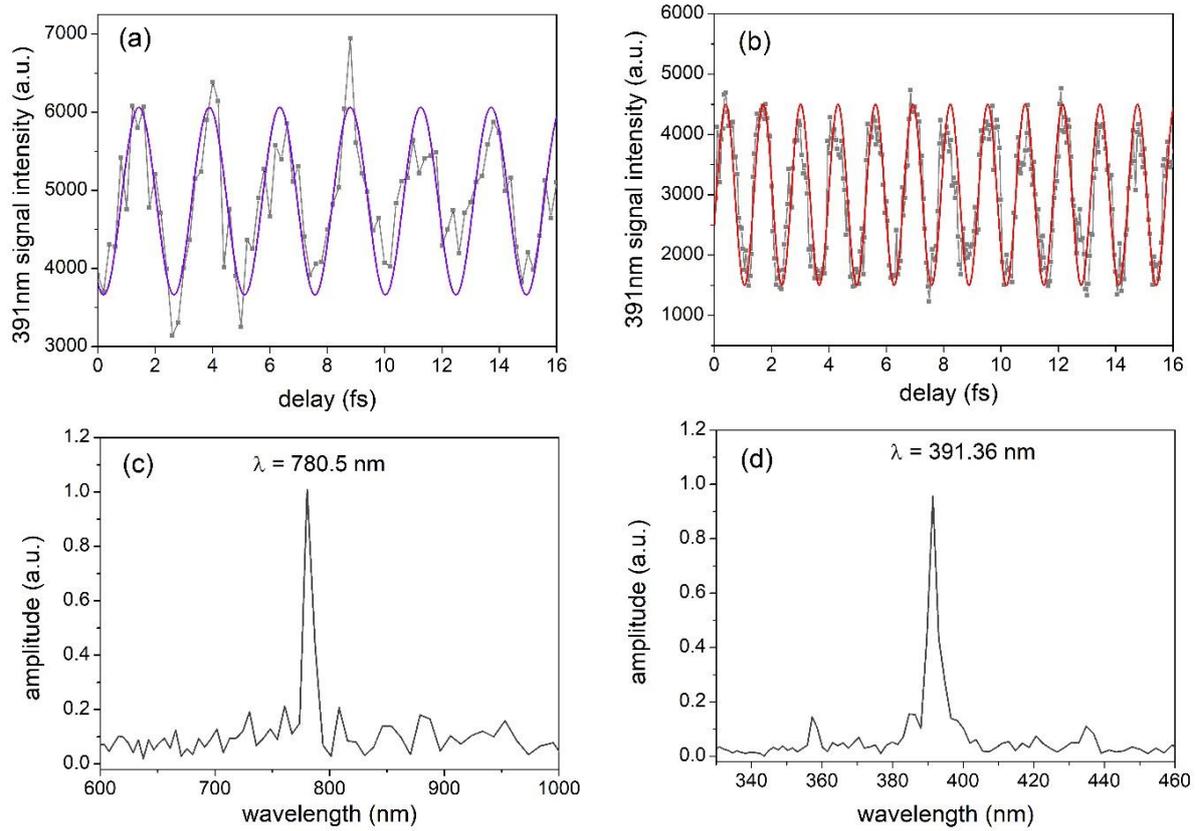

Fig. 3. (a): Measurement of the emission at 391.4 nm as a function of fine tuning of the delay between two laser pulses at 800 nm separated by 3 ps. (b): Measurement of the emission at 391.4 nm as a function of fine tuning of the delay between two seed pulses at 391 nm separated by 3 ps. The first seed pulse is delayed by 100 fs from the pump pulse. (c), Fourier transform of the modulation pattern of Figure 3 (a). (d) Fourier transform of Figure 3 (b).

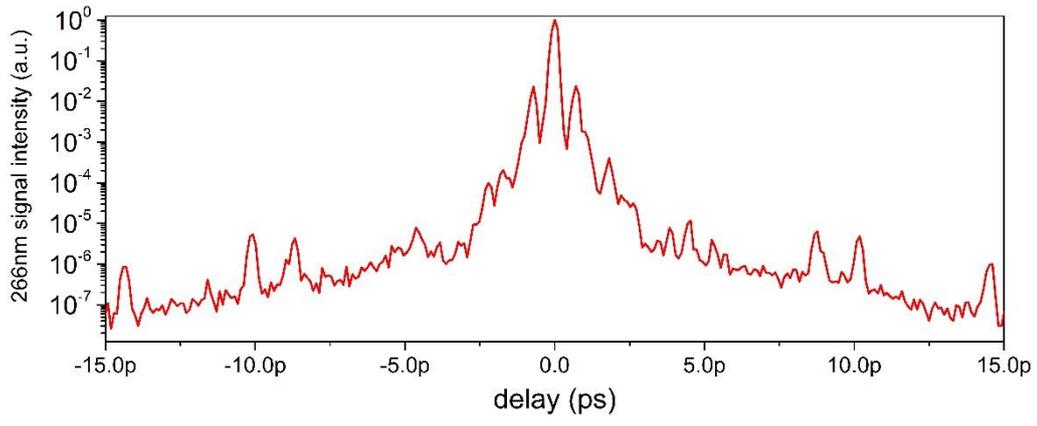

Fig. 4. Measurement of the contrast of the femtosecond laser system employed with a third order cross correlator (Sequoia, Amplitude Ltd.)

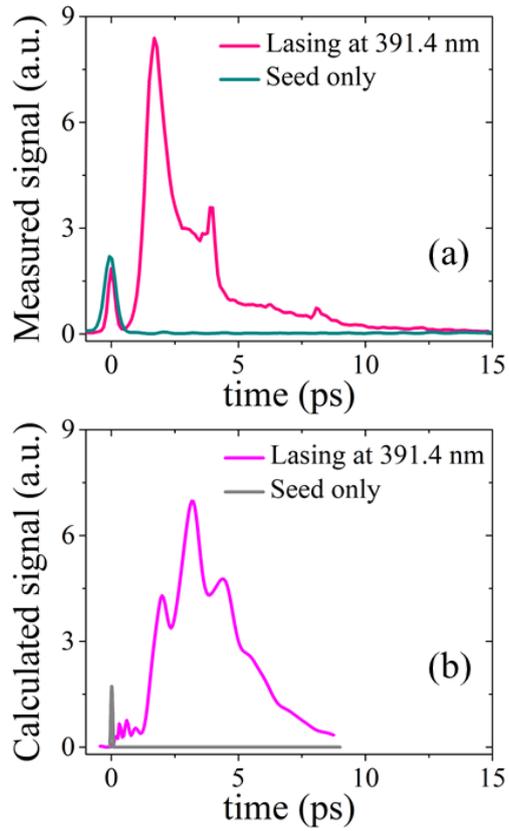

Fig. 5. Comparison between amplified seed signal forms obtained from experiment (a) and simulation (b) for nitrogen gas pressure $p$ = 30 mbar and a plasma length of 8 mm. Simulations are performed with following parameters: gas ionization 10%, initial populations at levels A and B: 10% and 30%, respectively, pump fluence: 20 mJ/cm$^2$, post pulse decay time 5 ps. Time $t$ = 0 corresponds to the end of the main pulse.

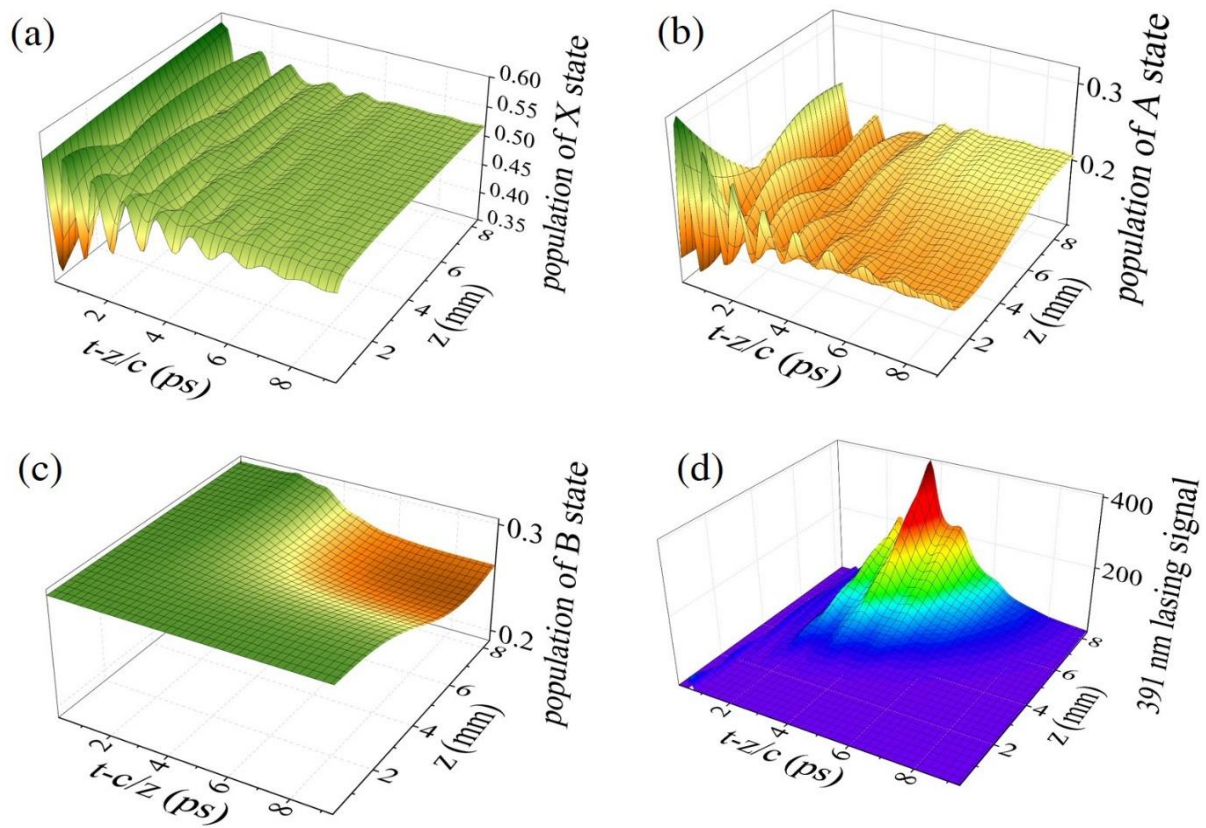

Fig. 6. Calculated evolution of populations in X, A, B and intensity of signal $I_{bx}$ in function of co-propagation time and plasma distance for a nitrogen pressure of 30 mbar. The populations are imposed at $t = 0$. Other parameters are as in Fig. 4 in the main text.

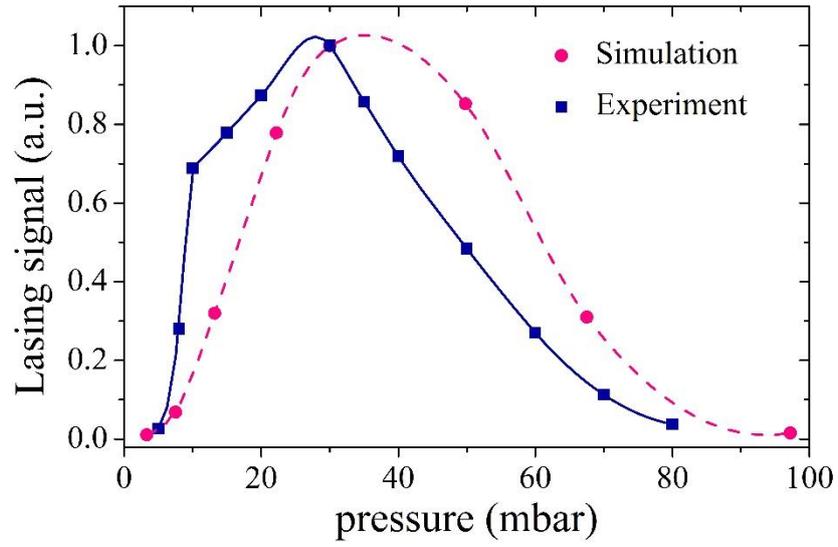

Fig. 7. Comparison between amplified seed signal obtained from experiments and simulations as a function of gas pressure. Simulation parameters are the same as in Fig. 5. The pump laser pulse with 2 mJ energy and 40 fs duration is focused with a lens of focal distance $f = 25$ cm.